\documentclass[11pt]{article}
\usepackage{moriond,epsfig}

\def\ApJ{Astrophys. J.}
\def\MNRAS{Month. Not. Roy. Ast. Soc.}

\def\AA{Astron. Astrophys.}

\def\be{\begin{equation}}
\def\ee{\end{equation}}
\def\bea{\begin{eqnarray}}
\def\eea{\end{eqnarray}}

\begin{document}
\vspace*{4cm}
\title{Archeops' results on the Cosmic Microwave Background}

\author{S. Henrot-Versill\'e (on behalf of the Archeops collaboration)}

\address{Laboratoire de l'acc\'el\'erateur lin\'eaire Orsay France}

\maketitle\abstracts{  Archeops is a balloon--borne experiment 
  dedicated to the measurement
  of the temperature anisotropies of the
  cosmic microwave background (CMB) from large angular scales to about 10 arcminutes. 
  A brief introduction to the CMB is given below,
  followed by a description of the Archeops
  experiment.
  Archeops flew on the 
  7th of February 2002 in the Arctic
  night from Kiruna (Sweden) to Russia. 
  The analysis of part of these data is described below with the
  results on the $C_\ell$ spectrum, showing for
  the first
  time a continuous
  link between the large scales and the first acoustic peak. 
  We end up with constraints on the cosmological parameters.
  We confirm the flatness of the Universe. And, combining the Archeops
  data with  other CMB experiments data and with the HST measurement of
	$H_0$, we measure
  for the first time $\Omega_{\Lambda}$ independently of 
  SuperNovae based results.}

\section{Brief introduction on the Cosmic Microwave Background}
At the very beginning of the history, the Universe was a
very hot soup of particles. As it expanded,
it became cooler and less dense. 

When the Universe was about 300.000 years old, it
was cold enough for the 
electrons and protons to combine and form hydrogen:
this is the time of recombination. At about the same
period, the photons decoupled from matter and the
Universe became transparent. These photons travelled
throughout the ages with almost no interaction with
the matter (except for the reionization period) and 
we can now detect them as the cosmic microwave background
or CMB.

At the time of recombination, the Universe had a different structure
that the one we know now:
matter was spread out very evenly. However
they should have been some structure in the early universe
to give birth to the galaxies and large scale structures 
we observe now.
And such slight increases in matter density
should have left an imprint on the CMB. 
Therefore when the photons decoupled from matter, they carried with them
the information on the matter distribution in the early universe.
Mapping the temperature fluctuations of these
photons as they appear to us now is mapping the matter
density
fluctuations as they were at a very early time,
when the photons last interacted with matter.

\subsection{The observable}

The goal of CMB experiments is then to  measure these 
temperature anisotropies (${\delta T \over T}$)
over  all the directions ($\theta$) of the sky, building
a map (or part of a map) of the Universe. 
Once we have obtained this map, we 
decompose it in terms
of spherical harmonics:
\be
{\delta T \over T } (\theta,\phi) = \sum_{l=0}^{\infty} \sum_{m=-l}^{l} a_{lm} Y_{lm}(\theta,\phi)
\ee
and we can get the power spectrum $C_\ell$ defined as:
\be
C_\ell={1\over 2l+1} \sum_{m=-l}^{l}|a_{lm}|^2
\ee
which contain all the information on the fluctuations
if they are Gaussian.
$\ell$ can be seen as the inverse of the size of the structures
on the sky: $\theta=1^\circ$ correspond to $\ell=200$.

\begin{figure}[h]
\begin{center}
\mbox{\epsfig{file=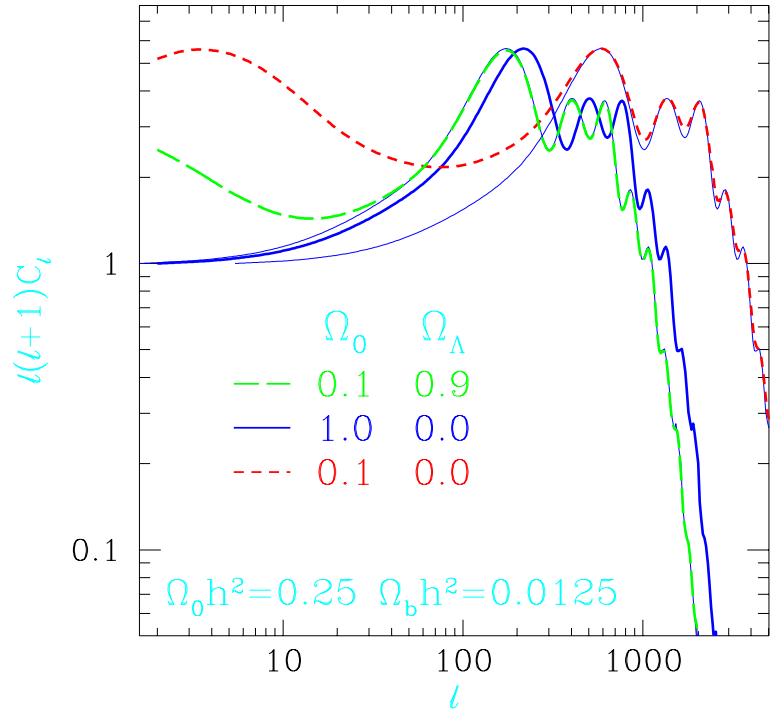,width=10cm,angle=270}}
\caption{\label{fig::whu} {Angular power spectrum for different
sets of cosmological parameters values: here $\Omega_{\Lambda}$ and $\Omega_0$
are being varied (from Wayne Hu web site).}}
\end{center}
\end{figure}

To extract the parameters which describe the early Universe,
we compare the
measured $C_\ell$ spectrum and 
the predictions  as it is shown on figure \ref{fig::whu} for
two varying parameters $\Omega_{\Lambda}$ and $\Omega_0$. We can then fit
for cosmological parameters such as the matter
density, the baryon density, the age of the Universe...
For instance the position
of the first acoustic peak is
highly related to the total density $\Omega_{\hbox{tot}}$.

\subsection{What we have learned so far}

The search for anisotropies of the CMB temperature across the sky began with Penzias and Wilson in 1965. 
They estimated the temperature to be isotropic to within about 10$\%$ (Penzias A. and Wilson R. \cite{PEtW}). 

In 1976, a flying instrument on a U2 spy plane established a 3mK dipolar temperature variation across the sky, 
arising from the motion of the Solar System with respect to the rest frame defined by the CMB. 

In 1989, NASA launched the Cosmic Background Explorer (COBE) (Mather J. et al \cite{COBE}), a satellite devoted to the study the microwave and infra-red backgrounds. The Far InfraRed Absolute Spectrometer (FIRAS) determined the CMB temperature
to be 2.728$\pm$ 0.002K, and showed that any spectral deviations from a Planck spectrum were less than 0.005$\%$. The CMB is the most precise known black body  and could only have arisen from the very hot, dense conditions that existed in the early Universe. COBE refined the dipole measurement showing that the Solar System velocity was 371$\pm$0.5km/sec in that frame. 

Since then, almost all the measurements on the CMB concentrated on smaller angular scales. 
Anisotropies have now been observed on small angular scales as it is shown 
on figure \ref{fig::figure_cl}.

The Boomerang collaboration, which used a balloon experiment with bolometer at 300mK
flying over Antarctica, provided a detailed map of the first peak which, besides falling at the 1-degree size predicted by inflation, also determined that the universe was flat.

\section{Archeops}
\subsection{The instrument}
 
The heart of the 
Archeops instrument is made of spider web bolometers
cooled down to
100mK using an open $^3$He--$^4$He dilution
cryostat. For each bolometers (21 in total in 4 frequency bands: 143, 217,
353 and 545 GHz), we have individual optics with horns and filters
at different temperature stages (0.1, 1.6, and 10K).

Before being measured by the bolometers, 
the photons from the Big Bang are first collected 
on a Gregorian off--axis aluminium telescope which provides
an angular resolution of about 8~arc-minutes at 143~GHz.
The technology is the same as for the Planck-HFI satellite.

The scanning strategy is to make circles on the sky during the
arctic night in order to minimise the background from the Sun.
The speed rate is of 2 round per minutes at an elevation of
$41^\circ$.

The goals of Archeops are twofold:
\begin{itemize}
\item{} On the scientific side: Thanks to the large sky coverage
allowed by the scanning strategy, the first goal of Archeops was
to link the plateau at low $\ell$ measured by COBE to the first  
acoustic peak measurements of balloon and on-ground experiments
(see Fig.~\ref{fig::figure_cl}). The second scientific goal is the measurement
of the polarisation of the galactic dust at 353GHz which 
is not described here.
\item{} On the technical side: Archeops is a test-bench for
Planck-HFI since we are using the same open cycle dilution to cool
down the bolometers to 100mK,
the same cold optics and the same spider web bolometers. Planck's
launch is planned for February 2007.
\end{itemize}

\subsection{Results on the power spectrum}
The results presented here correspond to 12 hours of data on two bolometers
(one at 143 and one at 217 GHz). 
We are only using for the $C_\ell$ computation
the north of the galactic plane.
This means analysing 8 millions of
data points for each bolometers.  

The measured noise is 
better than what was expected according to the Planck design:
of the order of 100$mK/\sqrt{Hz}$ for the two bolometers used in
the analysis presented here.

\begin{figure}[!ht]
\resizebox{\hsize}{10cm}{\includegraphics[clip]{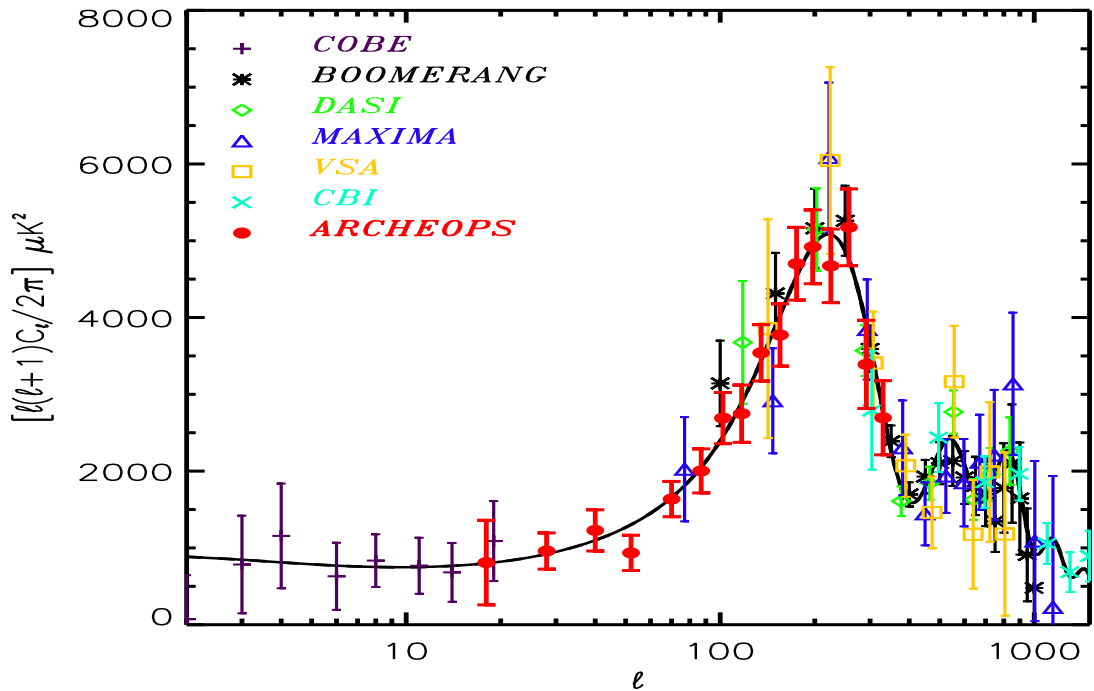}}
\caption{\label{fig::figure_cl}Archeops power spectrum in 16 bins along with some other
        recent experiments. A best model fit (continuous line) is
        obtained. The fitting allowed the gain of each experiment to
        vary within their quoted absolute uncertainties.
        Re-calibration factors, in temperature, which are applied in
        this figure, are 1.00, 0.96, 0.99, 1.00, 0.99, 1.00, and 1.01,
        for Cobe, Boomerang, Dasi, Maxima,
        Vsa, Cbi and Archeops resp., well within 1~$\sigma$ of the
        quoted absolute uncertainties ($<$ 1, 10, 4, 4, 3.5, 5 and
        7$\%$).}
\end{figure}

The Archeops 
$C_\ell$ spectrum is shown in red on 
Fig.~\ref{fig::figure_cl}
in 16 bins ranging from $\ell=15$ to $\ell=350$.
We also show in comparison a selection of other recent
experiments and a best--fit theoretical model. 

In order to study the systematic effects 
that could affect the results, we have made consistency checks
as for example a test of rotation (analysing the signal of
one circle and subtracting the one of the previous 
circle), and a difference test (analysing the time-line
of the signal at 143 subtracted by the one at 217 GHz).
For those two tests we have checked that the corresponding
power spectrum is compatible with 0 for all $\ell$.

In addition we have estimated the effect of the dust contamination
(mainly present at low $\ell$) and the bolometer time
constant and beam uncertainties (resp. at high $\ell$).
The have been found to be negligible with
respect to statistical error bars. The sample variance at low $\ell$
and the photon noise at high $\ell$ are the major contributors
to the final Archeops error bars of Fig.~\ref{fig::figure_cl}. 

As one can see on Fig.~\ref{fig::figure_cl}, the main goal
of Archeops (ie to provide an accurate link between
the large angular scales from Cobe and the first acoustic peak as
measured by degree--scale experiments like { Boomerang (de Benardis et al \cite{boom1}, Netterfield et al \cite{boom2})}, { Cbi (Sievers et al \cite{cbi})},
{ Dasi (Pryke et al \cite{dasi}) }, { Maxima (Hanany et al \cite{maxima1}, Netterfield et al \cite{maxima2})}, { Vsa (Rubi{\~n}o-Martin \cite{vsa})}) has been achieved.  
In addition we provide the best measurement of the first peak
before WMAP with a resolution of $\ell_{peak}=220\pm6$.

We have compared the map with the ones of Maxima (Hanany et al \cite{maxima1}, Lee et al \cite{maxima2})
and the ones of WMAP (Bennet et al \cite{wmap}), showing that
the same fluctuations are observed on all these maps
with a high correlation factor. 

\subsection{The cosmological parameters}

Using a large grid of
cosmological adiabatic inflationary models described by 7 parameters,
one can compute their likelihood with respect to the datasets. An
analysis of Archeops data only leads to put constraints on the total mass and energy density
of the Universe ($\Omega_{\hbox{tot}}$) to be greater than 0.9. Adding the
constraint of the measurement of $H_0$ by the HST (Freedman et al \cite{freedman}) we end up 
with the measurement $\Omega_{\hbox{tot}}=0.96^{+0.09}_{-0.04}$.

In combination with other CMB datasets (COBE, Dasi, Maxima, VSA, CBI) the Archeops data
constrain $\Omega_{\hbox{tot}}=1.15^{+0.12}_{-0.17}$ and the
spectral index $n=1.04^{+0.10}_{-0.12}$. In addition 
the baryon content of the Universe is measured to $\Omega_b h^2=
0.022^{+0.003}_{-0.004}$ which is compatible with the Big-Bang
nucleosynthesis  (O'Meara et al \cite{omeara}) and with a similar accuracy.

\begin{figure}[!ht]
\resizebox{\hsize}{12cm}{\includegraphics[clip]{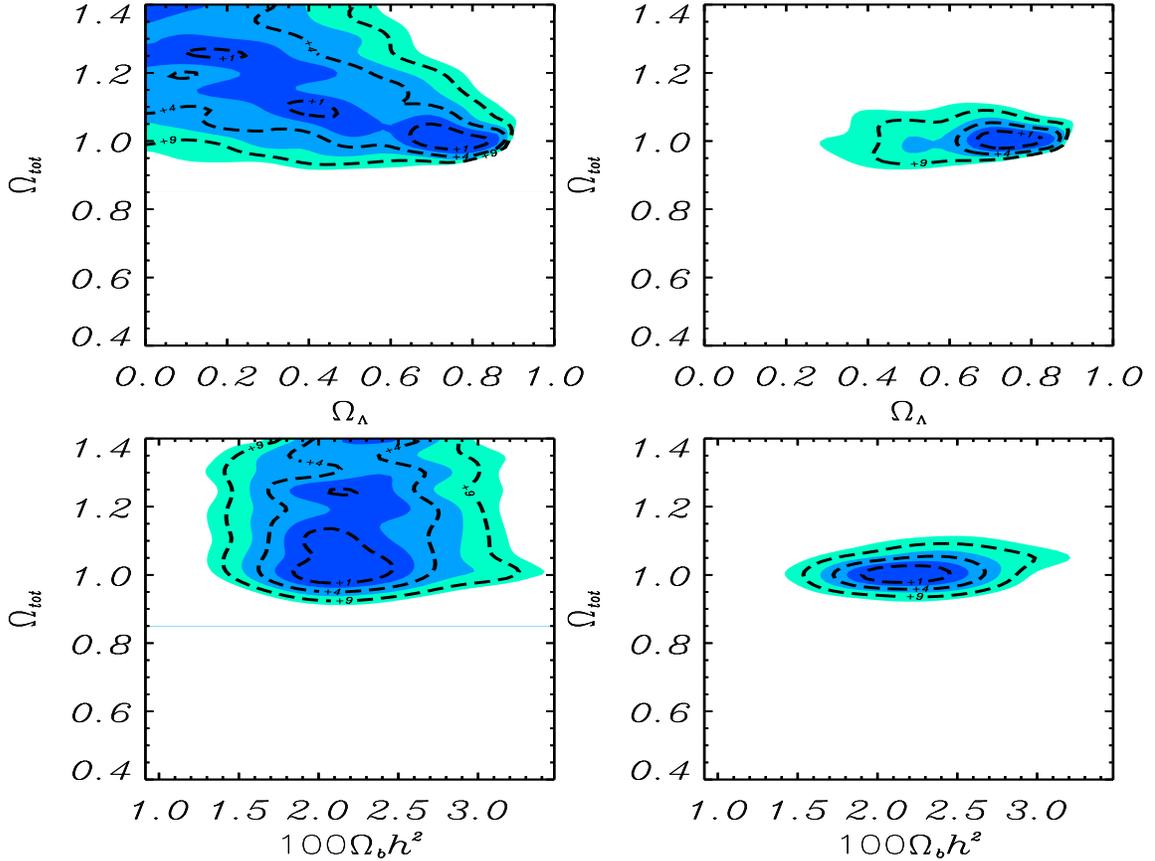}}
\caption{\label{fig::alllow}Likelihood contours in the $(\Omega_{\hbox{tot}}, \Omega_{\Lambda})$ and  
 $(\Omega_{\hbox{tot}}, \Omega_b h^2)$ planes. Left: constraints using Archeops and COBE, Dasy, Maxima, 
Boomerang, VSA, CBI
 datasets. Right: adding HST prior for $H_0$.}
\end{figure}

Using the recent HST determination of the Hubble constant (Freedman et al \cite{freedman}) 
leads to tight constraints on the total density, {\it
  e.g.}  $\Omega_{\hbox{tot}} =1.00^{+0.03}_{-0.02}$, ie the Universe is flat,
and permits to measure  $\Omega_\Lambda$
in an independent but compatible way with SuperNovae analysis:
$\Omega_\Lambda=0.73^{+0.09}_{-0.07}$.

The constraints are shown in the $(\Omega_{\hbox{tot}}, \Omega_{\Lambda})$ and  
 $(\Omega_{\hbox{tot}}, \Omega_b h^2)$ planes 
on figure \ref{fig::alllow}
without (on the left)
and with (on the right) the HST prior on the $H_0$ measurement.

\section{Summary}

For the first time we were able to fill the gap between
the large scales measured by COBE and the first acoustic
peak.

Combining Archeops measurements with all the CMB experiments
(before WMAP) we confirm that the Universe is flat and
combined with the $H_0$ measurement done by the HST,
we re-measure $\Omega_\Lambda$ independently of the SN based
results.

For Planck-HFI: the bolometer noise is better than the 
Planck design, and the open cycle dilution worked perfectly.

The analysis is in progress to measure galactic dust emission 
polarisation
with the Archeops last flight data. The use of all available bolometers
and of a larger sky fraction should yield an even more accurate and
broader CMB power spectrum in the near future. The large experience
gained on this balloon--borne experiment is providing a large feedback
to the Planck -- HFI data processing community.

For more details, two articles are available on Archeops: 
Beno{\^\i}t et al \cite{ar1}, 
and
Beno{\^\i}t et al 
\cite{ar2}.


\section*{References}

\begin{thebibliography}{99}
\bibitem{PEtW}
Penzias A. and Wilson R.: 1965, A measurement of excess Antenna Temperature at 4080Mc/s, \ApJ \it{Lett} {\bf 142} 419
\bibitem{COBE}
Mather J. et al: 1994, Measurement of the Cosmic Background Spectrum by the COBE-FIRAS Instrument, \ApJ {\bf{420}} 439 and
Smoot G. et al.: 1992, Structures in the COBE differential Microwave Radiometer First year Maps, \ApJ \it{Lett} {\bf 396} L1
\bibitem{maxima1}
Hanany, S.~et al,  \ApJ, {\bf 545}, L5, 2000
\bibitem{maxima2}
Lee, A. T., Ade, P., Balbi, A.~et al, \ApJ, {\bf 561}, L1-L6, 2001 
\bibitem{wmap} Bennet, C. L {\it et 
al.\/} 2003, astro-ph/0302207
\bibitem{boom1}
de Bernardis, P.~et al, Nature, {\bf 404}, 995, 2000
\bibitem{boom2}
Netterfield, C. B.~ et al, \ApJ, {\bf 571}, 604, 2002
\bibitem{cbi}
Sievers, J.~L.~et al, \ApJ, submitted, 2002, {\tt astro-ph/0205387}.
\bibitem{dasi}
Pryke, C.~et al, \ApJ, {\bf 568}, 46, 2002 
\bibitem{vsa}
Rubi{\~n}o-Martin~, J. A.~et al, \MNRAS, submitted, 2002, {\tt astro-ph/0205367}.

\bibitem{omeara}
O'Meara, J.~M.; Tytler, D., Kirkman, D., et al, \ApJ, {\bf 552}, 718, 2001
\bibitem{freedman} 
Freedman, W.~L., Madore, B.~F., Gibson, B.~K., et al, \ApJ,
  {\bf 553}, 47, 2001
\bibitem{ar1} Beno{\^\i}t  A. et al, 2003a,
  \AA, 399, L19,  {\tt astro-ph/0210305}
\bibitem{ar2}
Beno{\^\i}t A. et al,  2003b, \AA, 399, L25,
  {\tt astro-ph/0210306}

\end{thebibliography}
\end{document}